\documentclass{JINST}

\usepackage{graphicx}
\usepackage{amsmath}
\usepackage[tight,footnotesize, FIGTOPCAP]{subfigure}

\title{Performance of the Gas Gain Monitoring system of the CMS RPC muon detector and effective working point fine tuning}

\author{
S.~Colafranceschi$^b$\thanks{Corresponding author}, 
L.~Benussi$^a$, 
S.~Bianco$^a$, 
L.~Passamonti$^a$, 
D.~Piccolo$^a$, 
D.~Pierluigi$^a$, 
A.~Russo$^a$, 
G.~Saviano$^c$, 
C.~Vendittozzi$^c$,
M.~Abbrescia$^d$, 
A.~Aleksandrov$^e$, 
U.~Berzano$^b$, 
C.~Calabria$^e$,
C.~Carrillo$^b$,
A.~Colaleo$^d$, 
V.~Genchev$^e$, 
P.~Iaydjiev$^e$,
M.~Kang$^e$,
K.~S.~Lee$^e$,
F.~Loddo$^d$, 
S.~K.~Park$^e$,
G.~Pugliese$^d$,
M.~Maggi$^d$,
S.~Shin$^e$,
M.~Rodozov$^e$, 
M.~Shopova,$^e$,
G.~Sultanov$^e$,
P.~Verwillingen$^e$,
Et al.\\
  \llap{$^a$}Laboratori Nazionali Frascati dell'INFN.\\ 
  \llap{$^b$}CERN, Laboratori Nazionali Frascati dell'INFN and Universit\`a degli studi di Roma - La Sapienza.\\ 
  \llap{$^c$}Laboratori Nazionali Frascati dell'INFN and Sapienza Universit\`a di Roma. 
  \llap{$^d$}Politecnico di Bari, Universit\`a di Bari, and INFN Sezione di Bari, Bari, Italy.\\
  \llap{$^e$} to be completed.\\
  E-mail: \email{stefano.colafranceschi@cern.ch}
}

\abstract{
The Gas Gain Monitoring (GGM) system of the Resistive Plate Chamber (RPC) muon detector
in the Compact Muon Solenoid (CMS) experiment provides fast and accurate determination of
the stability in the working point conditions due to gas mixture changes in the closed loop recirculation
system. In 2011 the GGM began to operate using a feedback algorithm to control the applied voltage, in order to keep the GGM response insensitive to environmental temperature and atmospheric pressure variations. Recent results are presented on the feedback method used and on alternative algorithms.
}

\begin{document}

\section{Introduction}


The muon system of the Compact Muon Solenoid\cite{Chatrchyan:2008aa}  (CMS) experiment, at the LHC pp collider of CERN, Geneva (Switzerland), uses three different detector technologies: Drift Tube Chambers (DT), Cathode Strip Chambers (CSC) and Resistive Plate Chambers\cite{Santonico:1981sc} (RPC). As for any gas-based particle detector, the response of the CMS RPC system is strictly correlated to environmental variables, to the ratio of the gas components, and to the presence of pollutants that can be produced inside the gaps during discharges.
The CMS RPCs are bakelite-based double-gap RPCs operated with a 95.2\% C$_2$H$_2$F$_4$ - 4.5\% Iso-C$_4$H$_{10}$ - 0.3\% SF$_6$ gas mixture with an approximate 40\% relative water vapor content.
By design, the RPC gas system runs in closed loop\cite{gassystem}, with a fresh injected amount of gas limited to only 10\%.
 Therefore, the absence of gas mixture contaminants must be guaranteed  by means of suitable filtering  and monitoring systems.
\par
The Gas Gain Monitoring (GGM)\cite{Abbrescia:2007mu},\cite{Benussi:2008vs}
has been designed to monitor online the working point of the CMS RPC detector\cite{purifiers},\cite{Colafranceschi:2010zz},\cite{Bianco2010},\cite{Benussi:2010yx}
 for changes due to gas mixture differences in the Closed Loop (CL) recirculation system among fresh, before filters and after filters gas mixture. The monitoring is performed by checking the stability of the anodic charges ratios between gas mixtures, in order to cancel out common effects due to environmental parameters (temperature, humidity, atmospheric pressure). The GGM was installed and commissioned in 2010 on the CMS detector and has been taking data since then\cite{Benussi:2008fp},\cite{Benussi:2010yw}.
\par
Recently a study was performed aimed to provide the system with a high-voltage supply feedback system which allows to compensate for temperature $T$ and atmospheric pressure $p$ variations. With such a feedback algorithm the GGM charge distributions do not depend on $p$ and $T$, thus allowing one to monitor any gas mixture change common to the three (fresh, before-purifiers, after-purifiers) gas types.
\par
In this paper the status and collected results during 2011 data-taking are presented and discussed along with preliminary results on the $T$ and $p$ feedback algorithm.

\section{The Gas Gain Monitoring system setup}
\begin{figure}[ht]
  \centering
  \includegraphics[width=9cm]{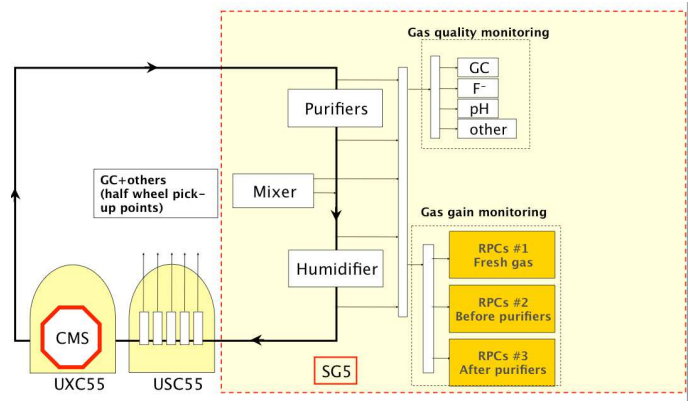}
  \caption{The GGM system integrated into CMS closed loop gas system.}
  \label{ggm_setup}
\end{figure}

The GGM (Fig.~\ref{ggm_setup}) is composed of twelve single-gap bakelite RPC detectors, with 2~mm thickness and   (50$\times$50) cm$^2$ area.
The setup is installed, on surface, in the CMS SGX5 building, close to CMS assembly hall, to profit from maximum cosmic muon rates necessary to provide a fast response.
%
%

The GGM consists of a cosmic-ray telescope of twelve RPC single gaps arranged in three sub-systems: "Fresh Gas", "Before purifiers" , "After purifiers".
The first sub-system (two gaps) is operated with the fresh CMS RPC gas mixture and is used as reference.
The second sub-system (three gaps) is operated with gas coming from the CMS RPC closed-loop gas system and extracted before the gas purifiers, while the third sub-system (three gaps) is operated with gas extracted after the gas purifiers.
The purpose of GGM is to monitor any deviation of the working point of the CMS RPC detector due to differences among the three gas types. This is accomplished by comparing the signal of cosmic muons in gaps flushed with different gas origin.
The GGM system runs continuously in a fully automatic way, each data sample consists in $10^4$ events that are collected every 30 minutes and the analysis is completed online providing to RPC operation a prompt working point measurement.

Each chamber of the GGM has a double side pad read-out, the signal is read-out by a transformer based circuit that allows to algebraically subtract the two signals, which have opposite polarities, and to obtain an output signal with subtraction of the coherent noise and with an improvement by about a factor 4 of the signal to noise ratio\cite{Benussi:2008vs}. The GGM RPC read-out pads are connected to a VME (VERSABUS Module Eurocard) ADC (Analog to Digital Converter) that is controlled by a semi-automatic DAQ system. 
%
%
All environmental parameters are continuously monitored (room temperature, relative humidity, atmospheric pressure).  Temperature, pressure and relative humidity sensors are also installed in the gas line before and after each chamber.
The accuracy of the temperature sensor is $\rm{\pm1^{\circ}C}$ in the range 0--40$^{\circ}$C and the resolution is 0.1$^{\circ}$C. The relative humidity sensor has an operating range from 2\% to 98\% with a 0.1\% resolution, $\pm$1\% absolute accuracy. The barometer operational range is between 700~mbar and 1050~mbar with a 0.1~mbar resolution and a $\pm$1~mbar accuracy.


\section{Feedback algorithm}
An HV (High Voltage) feedback function was added in 2011 to compensate for environmental conditions that affect the GGM chambers response. Such an algorithm has constituted a test ground for a possible application of a similar HV feedback to the full CMS RPC system. 
This solution aims to correct the applied voltage of each chamber maintaining its gain constant against environmental changes that would modify the working point of the chamber. 
The applied HV ($V$) is corrected according to the environmental pressure and temperature. The algorithm keeps stable the effective HV ($V_{eff}$) as in the Eq.~\ref{hveffi}~\cite{effective}:

\begin{equation}
\label{hveffi}
V_{eff} = V \cdot \frac{p_0  T}{p  T_0}
\end{equation} 

\noindent where p$_0$=965~mbar and T$_0$=293~K.

\begin{figure}[ht]
  \centering
  \includegraphics[width=7cm]{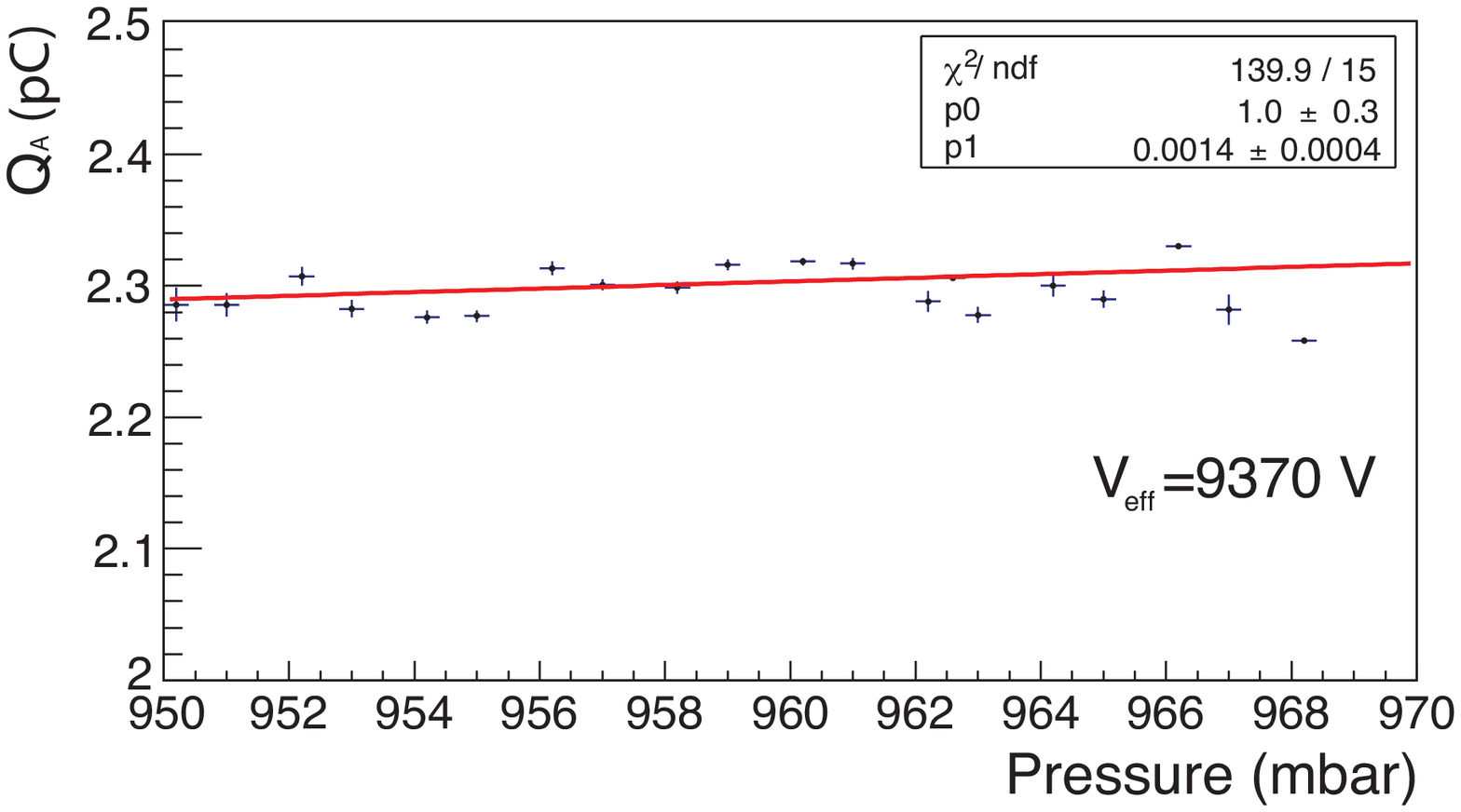}
  \put(-160,98){(a)}
  \includegraphics[width=7cm]{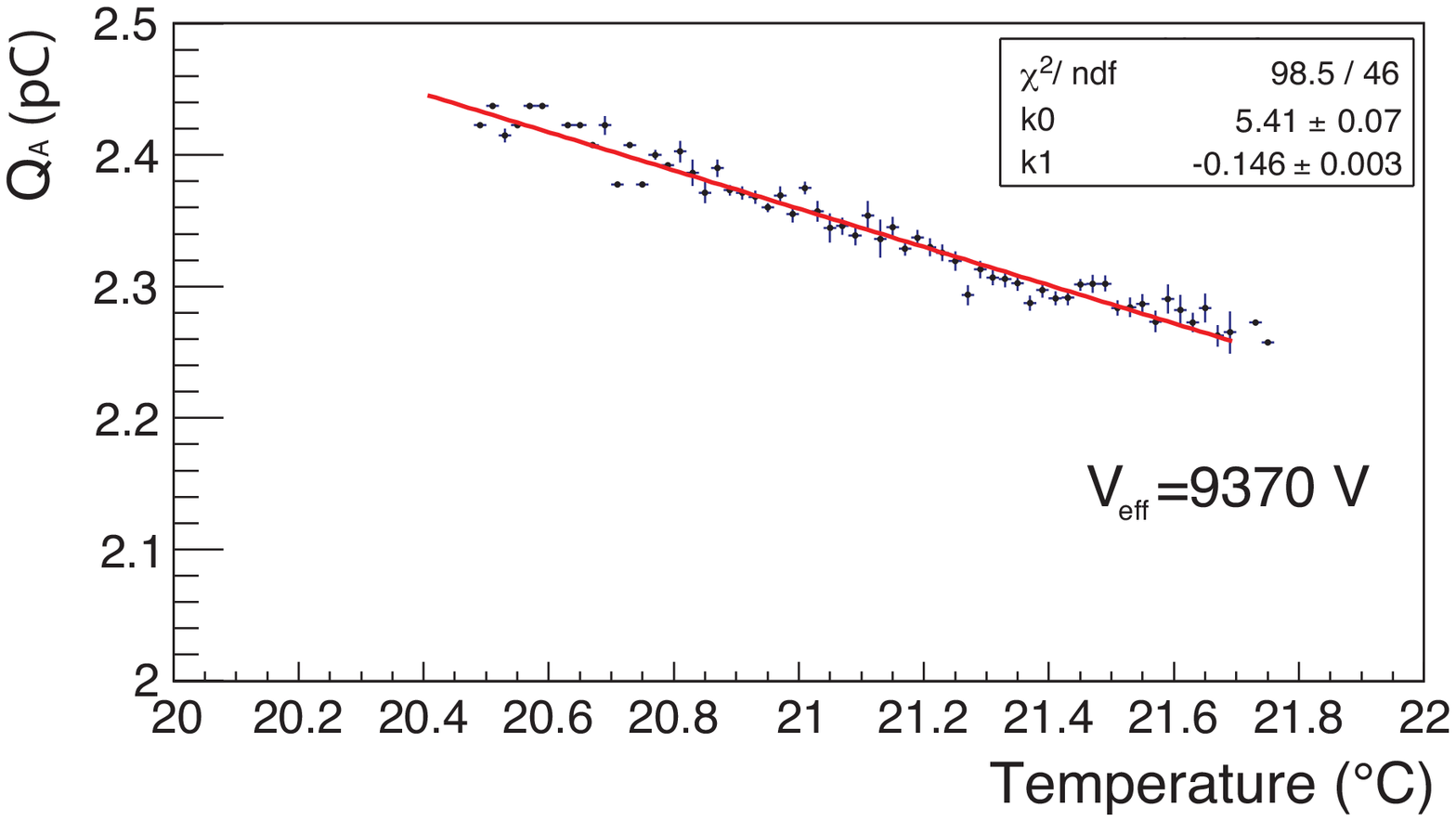}
  \put(-160,98){(b)}
  \hfil
%
%
  \caption{(a) Correlation plot between GGM RPC anodic charge against environmental pressure. (b) Correlation plot between GGM RPC anodic charge against environmental temperature. The meaning of fit parameters $p_0, p_1, k_0, k_1$ is described in the text.
}
  \label{ggm_fit}
\end{figure}

Figure~\ref{ggm_fit}(a) shows correlation plots between the anodic charge of one typical chamber against environmental pressure, while Fig.~\ref{ggm_fit}(b) shows the anodic charge against temperature for the same chamber. Both Fig.~\ref{ggm_fit}(a) and Fig.~\ref{ggm_fit}(b) refer to data taken with the chamber operated at the 50\% of efficiency, corresponding to an HV=9370~V. 
\par
 Data points in Fig.~\ref{ggm_fit}(a) and Fig.~\ref{ggm_fit}(b) have been fitted to linear functions:

\begin{equation}
\label{fitqa1}
Q_{A}(p) = p_{0} + p_{1}  p
\end{equation} 
\begin{equation}
\label{fitqa2}
Q_{A}(T) = k_{0} + k_{1}  T
\end{equation} 

\begin{figure}[h]
  \centering
  \includegraphics[width=7.5cm]{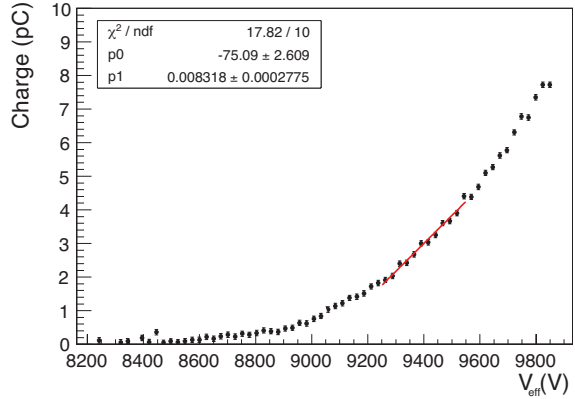}
  \label{chargetensione}
  \caption{Charge trend as a function of effective voltage one typical GGM chamber.}
\end{figure}

The fit in Fig.~\ref{ggm_fit}(a) returns very poor fit quality and parameters roughly compatible with a constant function. Work is ongoing to study the  residual systematics affecting the fit. The quality of fit in Fig.~\ref{ggm_fit}(b) is acceptable and the fit linear function describes adequately the dependence of $Q_a$ with respect  to atmospheric temperature. An overcorrection effect is evident. 
The feedback formula used (Eq.~\ref{hveffi}) overcorrects the applied HV by decreasing more than necessary the supply voltage. To characterize the overcorrection and determine a more suitable feedback correction function we have used formula in Eq.~\ref{hveffi2} from \cite{mbianco}:

\begin{equation}
\label{hveffi2}
V_{eff} = V \cdot \frac{p_0}{p} \cdot   \left[1+  \alpha_T \cdot \left(\frac{T}{T_0}-1\right)\right]
\end{equation} 

In order to determine the $\alpha_T$ parameter we use for each chamber its own gain curve (Q vs V curve). In Fig.~\ref{chargetensione} is shown a typical charge vs voltage dependence curve. We fit the data with a linear curve in a interval around the typical operation voltage. The interval is chosen in such a way to  take into account the total temperature excursion  over one year of operation at SGX5, expected to be about 3$^\circ$C corresponding to $\approx$90~V.
From Fig.~\ref{ggm_fit}(b)  and Fig.~\ref{chargetensione} the dependence between voltage and temperature corresponds to 17.5~V/$^\circ$C. Using the Eq.~\ref{hveffi2}, from the analysis of our data shown in Fig.~\ref{ggm_fit}(b) we find:

\begin{equation}
  \alpha_T = 0.40 \pm 0.05
\end{equation}


\section{Operational experience}
\begin{figure}[ht]
  \centering
  \includegraphics[width=9.8cm]{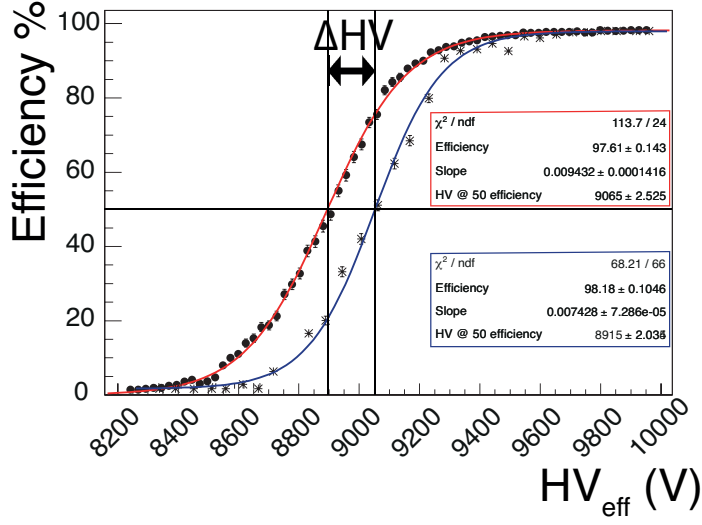}
  \includegraphics[width=7cm]{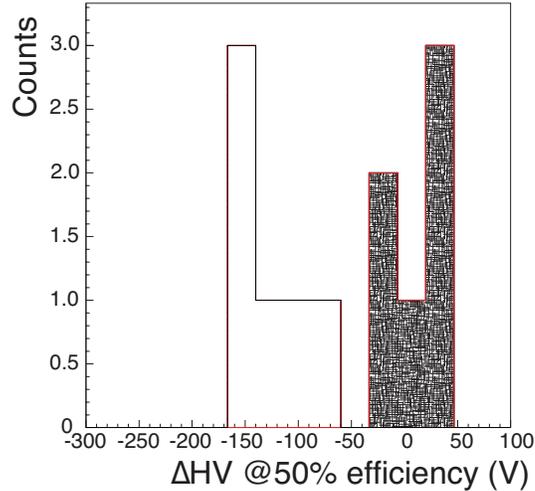}
  \label{hvscan}


  \caption{(top) Typical GGM high voltage scan performed during January 2010 (star) and April 2011 (full circle). (bottom) Difference in voltage between the HV scans performed in April 29th 2011 (correct gas mixture) with respect to and March 3th 2011 (empty histogram - wrong gas mixture) and January 10th 2010 (hatched histogram - correct gas mixture).}

\end{figure}

In March 2011 an hardware failure occurred to the CMS RPC gas Mass Flow Controller (MFC) leading to a wrong mixture injected into the closed loop. It was concluded that the faulty MFC was delivering about 34\% more SF$_6$ than designed. The content of SF$_6$ increased from 0.30\% to 0.34\% affecting the RPC working point.
The gain variation was shown by means of a series of HV scans which identified the faulty MFC (Fig.~\ref{hvscan}(top)  as example) and confirmed the presence of a wrong gas mixture. The function $\eta$ adopted to perform the HV efficiency scan fit is described by the Eq.~\ref{effi} from \cite{sigmoid}:

\begin{equation}
\label{effi}
\eta = \frac{\epsilon_{max}}{1+e^{-\lambda(V_{eff}-V_{50\%})}}
\end{equation} 

Fig.~\ref{hvscan}(bottom) shows the difference (in voltage) at the 50\% efficiency between HV scans performed with good and wrong gas mixture. The difference between April (good mixture) - March (wrong mixture) shows about 100~V difference at 50\% efficiency.

\section{Conclusions}
A feedback algorithm was developed and tested on the GGM system of the CMS RPC muon detector.
The feedback adjusts the HV supply in order to change the gain and therefore to stabilize the GGM working point against variations in atmospheric pressure and room temperature.
\par
Results show how the feedback function used (Eq.~\ref{hveffi}) provides an adequate correction for pressure but not for temperature. A study of the residual charge-temperature correlation was performed by using a modified feedback function (Eq.~\ref{hveffi2}), whose parameter was measured $\alpha_T = 0.40 \pm 0.05$. Our preliminary results are being finalized and will be used to correct  (and to be implemented in) the feedback function of the CMS RPC detector.
\par
Finally, the monitoring of the working point of the GGM performed via dedicated HV scans was shown to be a sensitive tool to assess the stability of the gas mixture, as shown for small variations of the SF$_6$ fraction occurred following a hardware failure of MFC.

\end{document}